\documentclass[graybox]{svmult}

\usepackage{type1cm}        
\usepackage{makeidx}         
\usepackage{graphicx}        
\usepackage{multicol}        
\usepackage[bottom]{footmisc}

\usepackage{newtxtext}       %
\usepackage[varvw]{newtxmath}       

\makeindex             

\begin{document}

\title*{Haloes, other dark matter candidates and astrophysical implications}
\author{Philippa S. Cole}
\institute{Philippa S. Cole \at Dipartimento di Fisica ``G. Occhialini'', Universit\'a degli Studi di Milano-Bicocca, Piazza della Scienza 3, 20126 Milano, Italy\\ INFN, Sezione di Milano-Bicocca, Piazza della Scienza 3, 20126 Milano, Italy\\ \email{philippa.cole@unimib.it}}

%
%
\maketitle

\abstract*{It is possible that a multi-component dark matter model is required if primordial black holes only contribute to a fraction of the energy density in dark matter. This is increasingly more likely with respect to the case of $f_{\rm PBH}=1$, since there is only one remaining window where primordial black holes can make up all of the dark matter on asteroid-mass scales. A mixed dark matter model can lead to interesting observables that come about due to the interactions between primordial black holes and the second dark matter component. This can provide unique signatures of the presence of primordial black holes and increase the prospects of detection or improvement of constraints in the mass ranges where $f_{\rm PBH}<1$.}

\abstract{It is possible that a multi-component dark matter model is required if primordial black holes only contribute to a fraction of the energy density in dark matter. This is increasingly more likely with respect to the case of $f_{\rm PBH}=1$, since there is only one remaining window, on asteroid-mass scales, where primordial black holes can make up all of the dark matter. A mixed dark matter model can lead to interesting observables that come about due to the interactions between primordial black holes and the second dark matter component. This can provide unique signatures of the presence of primordial black holes and increase the prospects of detection or improvement of constraints in the mass ranges where $f_{\rm PBH}<1$, whilst simultaneously exploring the remaining open parameter space for other dark matter candidates.}

\section{Introduction}
\label{sec:1}
Constraints on the abundance of PBHs have strengthened across the mass-range of PBHs that could be dark matter candidates, namely $10^{-18}\,\rm{M_\odot}\lesssim M_{\rm PBH}\lesssim10^{4}\,\rm{M_\odot}$, where $M_{\rm PBH}$ is PBH mass \cite{Green:2020jor,Carr:2021bzv}. There is only one window remaining around the mass of an asteroid, $10^{-16}\,\rm{M_\odot}\lesssim M_{\rm PBH}\lesssim10^{-10}\,\rm{M_\odot}$, where all of the energy density budget for dark matter could still be explained by PBHs of a single mass (or narrow distribution of masses) \cite{Montero_Camacho_2019}. See Part V of this volume for details of the current status of constraints. 

However, an order 1 - 10\% fraction of the dark matter can still be explained by PBHs in much of the parameter space, which would still provide a far greater understanding of the make up of dark matter than we currently have. Furthermore, it is important to stress that even in mass ranges where only a small fraction of the dark matter can be explained by PBHs, their existence would be enlightening for how the very first stages of the early universe evolved. This is because very specific ingredients are required from either inflationary mechanisms (see e.g. \cite{Ozsoy_2023,Ballesteros_2022,Mishra_2020,Cicoli_2018,Karam_2023,Hertzberg_2018,germani2017primordial,Bhaumik_2020} for a selection of examples) or cosmological processes such as phase transitions \cite{Lewicki_2023,Kawana_2022} in order to produce even just one PBH \cite{Cole:2017gle}. Producing a specific abundance of PBHs is an extremely delicate procedure \cite{Cole:2023wyx}, and needing to match an observed abundance today would be very prescriptive for identifying the correct model of the early universe. 

If it is the case that PBHs exist but only make up a fraction of the dark matter energy density budget, such that $f_{\rm PBH} = \frac{\rho_{\rm {PBH}}}{\rho_{\rm DM}}<1$, then there must be more than one species of dark matter. In this chapter, the other component (or components) of dark matter will be assumed to be of a particle nature. Whilst less simple, a `mixed' dark matter scenario offers opportunities for additional signatures owing to the interactions between PBHs and the other dark matter particles \cite{yang2023constraints,Boucenna_2018,PhysRevD.103.123532,10.1093/mnras/stad356,Kadota_2022,Adamek:2019gns}. This could provide additional observational probes, as well as new ways of constraining the abundance of both PBHs and particulate dark matter, which has also thus far eluded detection. From a cautionary perspective, the combined scenario could also lead to obscuring the existence of either species, not only because there would be a lower abundance of each, but also because current search strategies may not account for the presence of two species, that could result in signatures of dark matter being swept up in degeneracies \cite{Cole:2022ucw}.

\section{Particle companions}
\label{sec:2}
The two main classes of particle dark matter candidates that have been considered as companions to PBHs are weakly interacting massive particles (WIMPs) and axions, or axion-like particles (ALPs). We briefly introduce them in this section, before focusing on signatures of their possible co-existence with PBHs in the following sections.

\subsection{WIMPs}
Weakly interacting massive particles interact only via the weak force and gravity. They were postulated as a candidate that produced the correct abundance of dark matter to match cosmic microwave background observations with masses in the GeV to TeV range, i.e. at the electroweak scale, which is a natural energy scale at which to expect new physics \cite{STEIGMAN1985375, Giudice_2008,Feng_2013, Roszkowski_2018}. 

Briefly, dark matter particles could be produced as a thermal relic of the Big Bang such that during the early universe they would be in thermal equilibrium with all of the other particles. As the universe cools and expands, the dark matter particles would become sufficiently diffuse so that they can no longer annihilate. Their number density approaches a constant at this epoch, or `freezes out'. Theories that produce thermal relics predict a cross-section which depends on the mass of the dark matter particle and the weak-scale gauge coupling $g_{\rm weak}$ which is of order 1. For particle masses in the GeV - TeV range, the cross-section which is set by $m_X$ and $g_{\rm weak}$ automatically produces the correct abundance for today's dark matter energy density. This coincidence of correct abundance and energy scale is a phenomenon often dubbed the `WIMP miracle' \cite{Feng_2023,PhysRevD.33.1585}. 

\begin{figure}[b]
\sidecaption
\includegraphics[width = 10cm]{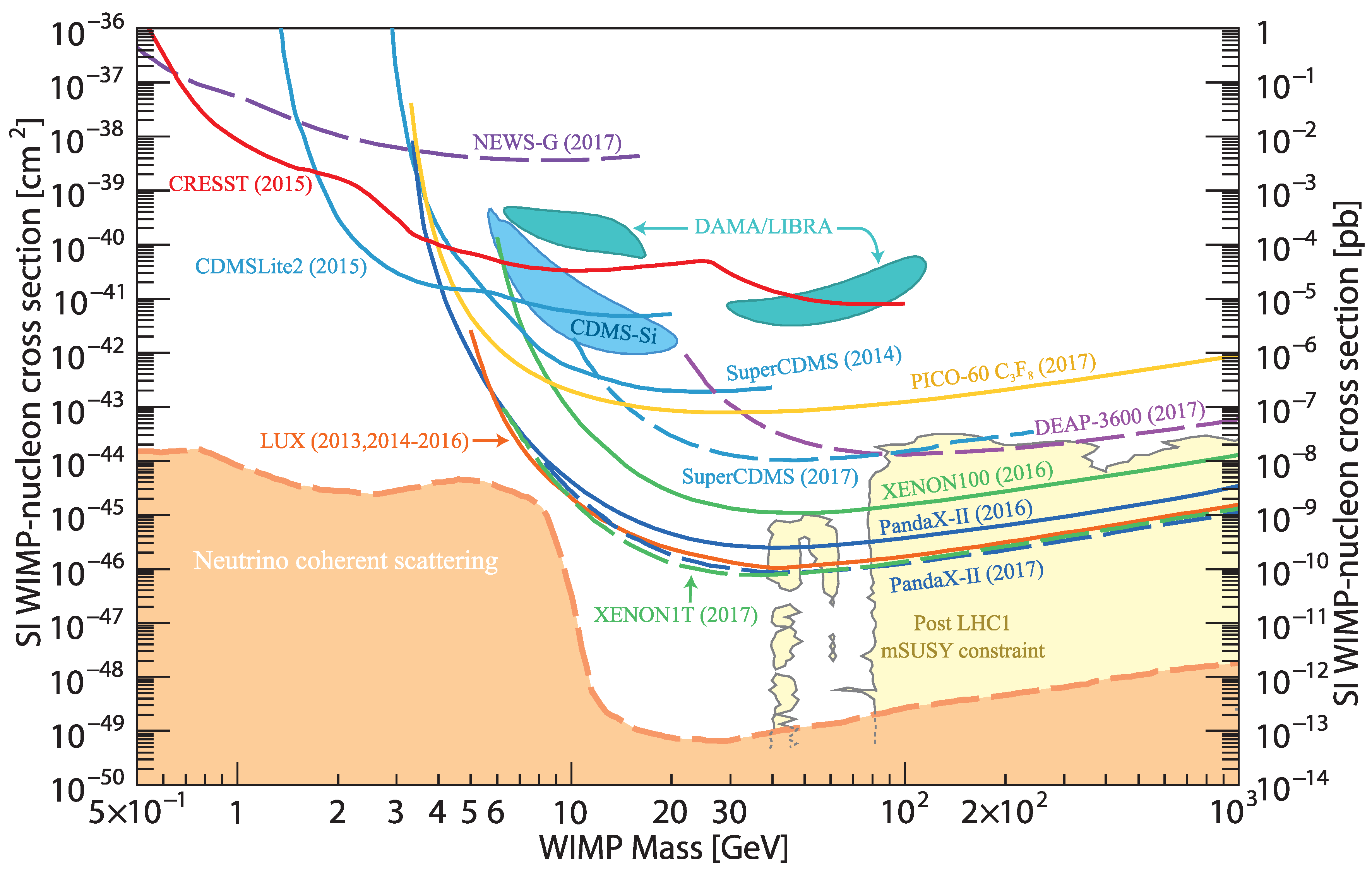}
%
%
\caption{Constraints on the WIMP cross section for spin-independent coupling as a function of WIMP mass. Reproduced under the Creative Commons by 4.0 license from \cite{chatterjee2019dark}.}
\label{fig:wimps}       
\end{figure}

Unfortunately, a plethora of searches for various WIMP models that employ different mediators for the interactions between the dark matter particles and the standard model have not found any evidence of their existence yet. For reviews, see \cite{Arcadi_2018,Schumann_2019}. Furthermore, there have been various critiques of the WIMP miracle \cite{Baer_2015}, decreasing their perceived novelty and broadening the space of viable dark matter candidates. Whilst this doesn't rule them out as dark matter candidates, they are certainly proving more difficult to find than first thought. Next generation experiments focusing on direct and indirect detection, as well as collider searches will continue to constrain the still-viable parameter space, see Fig.~\ref{fig:wimps}.\\

Nevertheless, the fact that WIMPs self-annihilate provides a unique observational signature for the combined scenario of WIMPs and PBHs, which ultimately leads to very tight constraints on their co-existence, see Sect.~\ref{sec:ann}.

\subsection{Axions and axion-like particles}
The quantum chromodynamics (QCD) axion was initially proposed as a particle that could explain the strong charge-parity (CP) problem \cite{Peccei_2008}. In a nutshell, the electric dipole moment of the neutron is constrained to be extremely small: $|d_n|<2.9\times10^{-26}e\,{\rm cm}$, where $e$ is the charge of the electron. In QCD, the theory of the strong interaction between quarks which is mediated by gluons, the value of $|d_n|$ depends on the mis-alignment angle $\theta_{\rm QCD}$ as
\begin{equation}
d_n \approx 3.6 \times10^{-16}\theta_{\rm QCD}e\,{\rm cm}.
\end{equation} 
Given the strong constraint on $|d_n|$, there is seemingly a fine-tuning problem on $\theta_{\rm QCD}\lesssim10^{-10}$, see e.g. \cite{Marsh_2016,Peccei_2008} for details. One solution to this fine-tuning problem is the QCD axion, which sets $\theta_{\rm QCD}=0$ via QCD non-perturbative effects. The QCD axion can be stable and have masses in the range $m_a\sim10^{-12} -10^{-3}\,{\rm eV}$, which means that it simultaneously qualifies as a viable dark matter candidate \cite{Marsh_2016}. 

\begin{figure}[b]
\sidecaption
\includegraphics[width = 10cm]{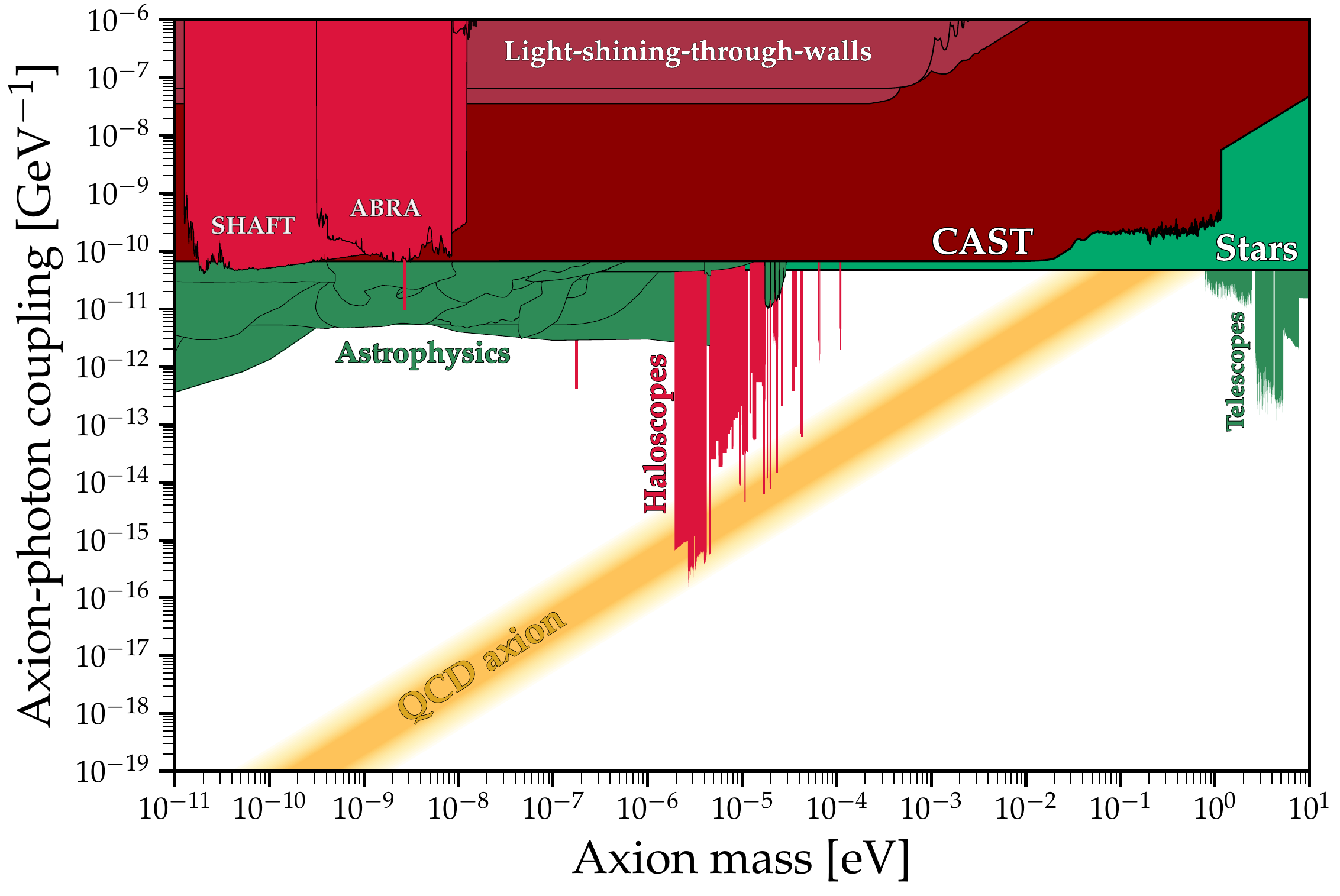}
%
%
\caption{Constraints on the axion-photon coupling as a function of axion mass. The yellow diagonal shaded region shows the region of the parameter space that corresponds to the QCD axion. Figure from \cite{AxionLimits} and further details in \cite{OHare:2024nmr}.}
\label{fig:axions}       
\end{figure}

However, many high energy physics models predict the existence of a more general class of axion-like particles (ALPs) and thus there is a much wider mass range of ALPs, $m_a\sim10^{-33}  - 10^{-3}\,{\rm eV}$ that can be considered as dark matter candidates \cite{Marsh:2017hbv}. As they are too light to be observed at particle accelerators, prospects for their detection tend to rely on looking for signatures of their conversion to and from photons, and their effect on astrophysical and cosmological observables. 
Targeted searches have found no evidence of them yet, including helioscopes such as the CERN Axion Solar Telescope (CAST) \cite{PhysRevLett.94.121301} which look for signatures of solar axions that have converted into x-rays, or haloscopes such as the Axion Dark Matter Experiment (ADMX) \cite{Khatiwada_2021} which aims to convert dark matter axions into microwave photons, as well as limits from astrophysical and cosmological data-sets such as CMB observations \cite{Bal_zs_2022}, globular cluster observations \cite{Dolan_2022} and white dwarf observations \cite{Dolan_2021}. However the parameter space, which is usually presented in terms of their mass and their coupling to the photon $g_{a\gamma}$, is vast, and there are various regions wide open in terms of constraints, see Fig.~\ref{fig:axions}.

Unique probes such as the signatures of ALPs co-existing with primordial black holes offer new ways of searching for them, and in this case, a distinct channel from other popular candidates such as WIMPs, as we will see in the following sections. 

\subsection{Other candidates}
There are many other proposals for dark matter candidates, including but not limited to self-interacting dark matter \cite{Tulin_2018}, warm dark matter \cite{Bode_2001}, or certain scenarios of modified gravity \cite{Milgrom_2015}. To date they have not been considered explicitly in combination with the co-existence of primordial black holes, however in Sect.~\ref{sec:evap}, we will consider the production of dark matter particles including some of these candidates from the evaporation of very light PBHs \cite{Bernal_2021}.

\section{Formation of dark matter haloes around PBHs}
\label{sec:halo}
If PBHs co-exist with particles of cold dark matter in the early universe, the gravitational attraction of the PBHs will overcome the decelerating expansion of the background during radiation domination. This causes PBHs to accrete the dark matter particles in the form of dense haloes. The characteristic time and radius of this process is at `turn-around', when and where the dark matter particles stop moving away from the PBH due to the background expansion and instead start falling inwards. Following the argument in \cite{Adamek:2019gns}, an analytical approximation identifies turn-around as when the Newtonian gravitational attraction of the PBH is equal to the deceleration of the background in radiation domination. The acceleration, denoted by double dots, of a point particle at a separation $r$ from a PBH is
\begin{equation}\label{eq:rddot}
	\ddot{r} = -\frac{GM_{\rm PBH}}{r^2} + \frac{\ddot{a}}{a}r
\end{equation}
where $G$ is Newton's gravitational constant, and $a$ is the scale factor. Then, equating the first and second term on the right hand side leads to the definition of the turn-around radius $r_{\rm TA}$:
\begin{equation}\label{eq:TA}
GM_{\rm PBH} = \frac{8\pi G}{3} \rho_{\rm rad} r_{\rm TA}^3
\end{equation}
where $\rho_{\rm rad}$ is the total background energy density during radiation domination and we have used the acceleration equation
\begin{equation}
\frac{\ddot{a}}{a} = -\frac{(1 + 3\omega)}{2}H^2
\end{equation}
with $\omega=1/3$ the equation of state during radiation domination and $H$ the Hubble factor, as well as the Friedmann equation in a flat universe
\begin{equation}
H^2 = \frac{8\pi G}{3}\rho.
\end{equation} 
Substituting the energy density at matter radiation equality $\rho_{\rm eq} = 2 \rho_{\rm rad}$ into Eq.~\ref{eq:TA}, demonstrates that the mass of the halo within the turn-around radius is approximately equal to the mass of the PBH at matter-radiation equality:
\begin{equation}
M_{\rm PBH} = \frac{4\pi}{3}\rho_{\rm eq}r_{\rm TA}^3.
\end{equation}

The dark matter density profile is then found by assuming that particles on radii smaller than the turn-around radius are frozen into the dark matter halo, and hence the dark matter energy density of the halo is equal to the background radiation energy density at turn-around time. Since matter scales like $a^{-3}$ in radiation domination, the dark matter density profile in terms of $\rho_{\rm eq}$ is
\begin{eqnarray}
\rho_{\rm DM}(r) &=& \rho(r(t_{\rm TA})) \nonumber \\
&=& \frac{\rho_{\rm eq}}{2}\left(\frac{a_{\rm TA}}{a_{\rm{eq}}}\right)^{-3} \nonumber \\
&=& \frac{\rho_{\rm eq}}{2}\left(\frac{t_{\rm TA}}{t_{\rm{eq}}}\right)^{-\frac{3}{2}} \nonumber \\
&=&  \frac{\rho_{\rm eq}}{2} t_{\rm eq}^\frac{3}{2} (8G M_{\rm PBH})^\frac{3}{4}r(t_{\rm TA})^{-\frac{9}{4}}
\end{eqnarray}
using $t=\frac{1}{2H}$ during radiation domination and the Friedmann equation again together with $\rho(t_{\rm TA}) \approx 3M_{\rm PBH}/4\pi r(t_{\rm TA})^3$. A numerical solution to the differential equation in Eq.~\ref{eq:rddot} shows that in fact the turn-around radius is more accurately identified at \cite{Adamek:2019gns}
\begin{equation}
GM_{\rm PBH} = 2H^2r(t_{\rm TA})^3
\end{equation}
i.e. with an additional factor of 2 with respect to Eq.~\ref{eq:TA}. Working this through results in a firm prediction for the dark matter density profile around PBHs of 
\begin{equation}\label{eq:profile}
\rho_{\rm DM}(r) = \frac{\rho_{\rm eq}}{2} t_{\rm eq}^\frac{3}{2} (2G M_{\rm PBH})^\frac{3}{4}r^{-\frac{9}{4}}
\end{equation}
where we have dropped the subscript $TA$ and it is implicit that the halo density is fixed at turn-around time.

\begin{figure}[b]
\sidecaption
\includegraphics[width=7.8cm]{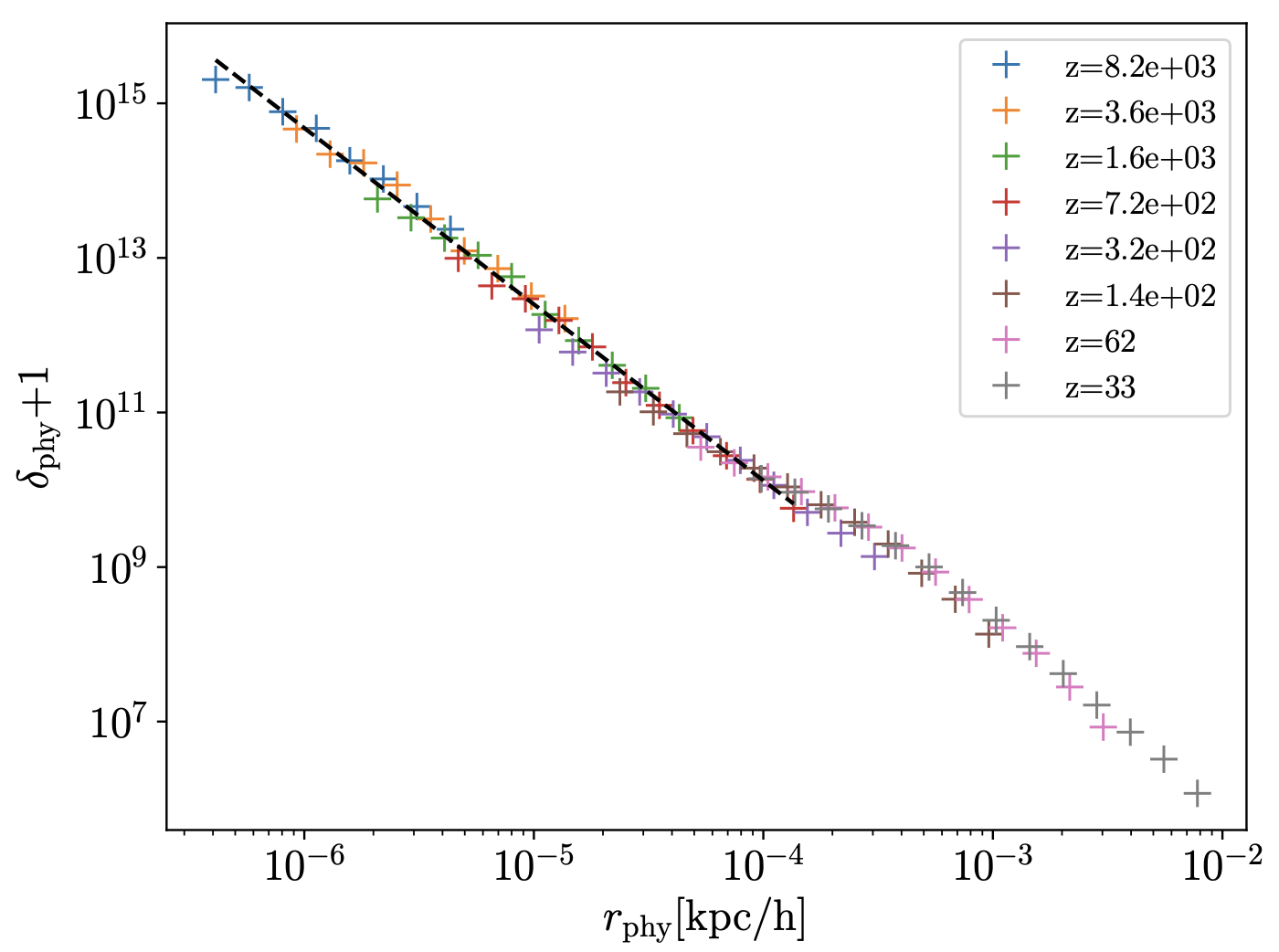}
%
%
\caption{The density profile as a function of radius of the dark matter halo surrounding a PBH. Coloured cross-markers are the results of simulations at the labelled redshift values, the dashed black line shows the analytical $\gamma_{\rm sp}=9/4$ prediction for the slope. Figure re-produced with permission from \cite{Adamek:2019gns}.}
\label{fig:profile}       
\end{figure}

Numerical simulations corroborate this analytical prediction \cite{Adamek:2019gns, Eroshenko:2016yve}, see Fig.~\ref{fig:profile}, for the dark matter density profile around $\mathcal{O}(10)\,{\rm M_\odot}$ black holes forming in radiation domination and evolved from redshift $z\sim8000$ down to redshifts between $z\sim30 - 60$. The results are expected to hold for any cold dark matter candidate, and should also be independent of the mass of the black hole, in the sense that a dense spike should form with a profile characterised by a slope of $9/4$ and a normalisation as given by Eq.~\ref{eq:profile}.
\\
\\

It is important to stress that if there exists a small population of PBHs accompanied by any cold dark matter particles, then dense haloes (also referred to as spikes) necessarily form around the black holes. Provided they are not significantly disrupted by, for example, close encounters or collisions with nearby stars or black holes \cite{Kavanagh:2018ggo}, then searches for one component must take into account the presence of the other. The next sections consider signatures to be expected from the co-existence of PBHs and a sub-set of particle dark matter candidates.

\subsection{WIMP haloes}
\label{sec:ann}

If the dense dark matter spikes around PBHs are made up of weakly interacting massive particles which interact amongst themselves with a given cross-section, then close to the black hole, the WIMPs will be packed close enough in that they will annihilate each other. This causes a plateau in the dark matter density profile at small radii, and also results in an annihilation signal that should be observable in gamma-rays.

The maximum density of WIMPs in the halo $\rho_{\rm max}$ depends on the mass $m_\chi$ and the cross-section $\langle\sigma v\rangle$ of the particles
\begin{equation}\label{eq:rhomax}
\rho_{\rm max} = \frac{m_\chi}{\langle\sigma v\rangle t_0}
\end{equation}
where $t_0$ is the age of the universe today. This means that there is a threshold radius of the dark matter profile, found by equating Eq.~\ref{eq:profile} and Eq.~\ref{eq:rhomax}, such that particles on smaller radii than the threshold will contribute to an annihilation signal, since the density of the spike exceeds the maximum value on these radii. 

Assuming that the annihilation signal contributes to the diffuse gamma-ray background, constraints on the decay rate of dark matter from \cite{Ando:2015qda} can be used to place constraints on $f_{\rm PBH}$ for $\mathcal{O}(10)\,{\rm M_\odot}$ PBHs. For the standard WIMP scenario with $m_\chi = 100\,{\rm GeV}$ and $\langle\sigma v\rangle = 3\times10^{-26}\,{\rm cm^3s^{-1}}$, $\mathcal{O}(10)\,{\rm M_\odot}$ PBHs are restricted to only making up a very small fraction of dark matter $f_{\rm PBH}\lesssim2\times10^{-9}$ \cite{Adamek:2019gns,Eroshenko:2016yve,Boucenna_2018}. Conversely, the detection of just one PBH would fix extremely strong constraints on almost all dark matter models motivated by weak-scale extensions of the standard model \cite{Bertone:2019vsk}.

This holds across the range of WIMP masses (approximately $m_\chi = 10 - 10^4\,{\rm GeV}$) and cross-sections in which there are constraints on the diffuse gamma-ray background (which depends on the decay channel assumed), and the constraint on $f_{\rm PBH}$ only varies by $\mathcal{O}(1)$ \cite{Ando:2015qda}. However, note that for much larger PBH masses, the number density of PBHs will be much smaller than for the case of $\mathcal{O}(10){\rm M\odot}$ PBHs and therefore the diffuse gamma-ray background constraints do not apply. Instead, the analysis should be done with point source observations from e.g. the Fermi satellite \cite{MacGibbon:2015mya}. Also for much lighter PBHs, the thermal kinetic energy of the WIMPs could alter the density profile of the dark matter depending on its mass. Therefore, the strong conclusion that WIMPs and PBHs effectively can't co-exist is only true for stellar-mass black holes and $m_\chi=10-10^4\,{\rm GeV}$. Mass ranges of either the WIMPs or the PBHs that differ greatly should be investigated independently. 

\subsection{ALP haloes and clouds}
\label{sec:alps}
Since axion-like particles do not self-annihilate, they are not subject to the same bounds from astrophysical observations that WIMPS are, as described in Sect.~\ref{sec:ann}. This means that across the full range of masses, PBHs could be accompanied by dense spikes of axion-like particles accumulated in the manner described in Sect.~\ref{sec:halo}, if they do not make up the entirety of the dark matter themselves. Possibilities for detecting such systems will be explored in Sect.~\ref{sec:searches}.

However, due to the very light masses of ALPs, there is another possibility for the formation of particle clouds around PBHs. If the primordial black holes are spinning, then an extremely dense boson cloud can form due to superradiance \cite{Lasenby_2016,Brito_2020}. This occurs when the angular momentum of the black hole is extracted and transferred to the ALPs if the Compton wavelength of the particles is comparable to the Schwarzschild radius of the black hole. The mass of the boson $m_b$ which has a comparable Compton wavelength to the Schwarzschild radius of a PBH with mass $m_{\rm PBH}$ is given by \cite{Franciolini_2022}
\begin{equation}
m_b \sim 10^{-4}\,{\rm eV}\left(\frac{m_{\rm PBH}}{10^{-6}\,{\rm M_\odot}}\right)^{-1}.
\end{equation}
Boson clouds tend to be much more dense than the spikes referred to in Sect.~\ref{sec:halo}, and therefore could provide very distinctive signatures that will be discussed further in Sect.~\ref{sec:searches}. 

However, in many PBH scenarios it is not expected that the PBHs should possess any significant spin \cite{Mirbabayi:2019uph}, unless they are formed during matter domination instead of radiation domination \cite{Harada:2017fjm}, or if they are formed not from the collapse of large scalar perturbations but instead from e.g. the collapse of domain walls. Even if superradiance took place during the early universe, the angular momentum of the PBH is expected to be depleted on a short time-scale and therefore its signatures may no longer be observable in the present epoch. Other possibilities for generating spin in the late universe include spinning up due to accretion \cite{Luca_2020}, or that the merger of two PBHs could form a spinning remnant \cite{Barausse_2009}.

\section{Searches for PBHs and dark matter with gravitational waves}
\label{sec:searches}
One of the most promising ways of probing the nature of dark matter, be it a particle or compact object, is via gravitational wave observations \cite{Bertone:2019irm}. Furthermore, there are distinct gravitational wave signatures for systems of primordial black holes surrounded by spikes or haloes of particle dark matter \cite{Kavanagh:2018ggo}. Improved sensitivities and wider frequency reaches of third generation detectors such as the proposed Einstein Telescope \cite{Punturo:2010zz} and Cosmic Explorer \cite{Evans:2021gyd} pose opportunities to simultaneously learn about two components of the dark matter. Either, through detecting PBH binaries merging in an environment of dark matter particles, or optimistically, through the signature of superradiance around a lone PBH.

\subsection{Dephasing signatures in the gravitational waveform}
As described in Sec.~\ref{sec:halo}, PBHs that do not make up all of the dark matter will necessarily have very dense dark matter particle haloes around them. If two PBHs form a binary then they will emit gravitational waves as they inspiral and eventually merge \cite{Nishikawa_2019}. If they are of similar mass, then it is likely that within the first few orbits the dark matter haloes will be disrupted and possibly completely destroyed \cite{Kavanagh:2018ggo}. This means that the binary system will behave in the same way as one in vacuum, and hence the gravitational wave signal will not display evidence of the dark matter particles. This is true for heavier particle-like dark matter, however it is unclear whether this assumptions holds for light, wave-like dark matter. Simulations have shown some evidence for the persistence of the cloud of dark matter in the presence of equal-mass binaries \cite{Aurrekoetxea:2023jwk}, however it remains to be seen whether this is a physical effect or a consequence of the boundary conditions within the simulations that do not allow dispersion of the dark matter beyond the extent of the halo.

However, if there is a sizeable difference between the masses of the two PBHs, such that the mass ratio $q=m_2/m_1\lesssim10^{-2.5}$ where $m_2$ is the mass of the lighter PBH, then the dark matter halo is likely to survive for the duration of the inspiral, and the gravitational waveform will be altered by the presence of the dark matter \cite{Kavanagh:2020cfn}. This provides an opportunity for signatures of dark matter to be detected by the merging of intermediate or extreme mass ratio PBH binaries.

\subsubsection{Dynamical friction}\label{sec:df}
As the companion (lighter) PBH drives through the dark matter halo of the primary PBH, the dark matter particles will form a wake behind the companion which slows down its velocity in the orbital direction, but increases its velocity in the radial direction. This means that the inspiral will complete in fewer cycles from a given entry frequency than an equivalent system in vacuum with the same masses. This effect is known as dynamical friction, and the difference in number of cycles is known as an accumulated dephasing of the gravitational waveform.

At zero-th Post-Newtonian order where the orbits of the companion object are approximated as quasi-circular, it enters into the waveform via the change in separation of the two black holes as \cite{Eda_2015,Kavanagh:2020cfn}
\begin{equation}
\dot{r}(t) = -\frac{64G^3Mm_1m_2}{5c^5r(t)^3} - \frac{8\pi G^\frac{1}{2}m_2\rho_{\rm DM}(r,t)\xi(r,t)r(t)^{\frac{5}{2}} \log\Lambda }{\sqrt{M}m_1}
\end{equation}
where the first term is the emission of gravitational waves, and the second is the Chandrasekhar formula for dynamical friction. $M = m_1 + m_2$, $\xi$ describes what fraction of the dark matter particles are moving more slowly than the companion object and can therefore be deflected and contribute to the drag force, $\Lambda=\sqrt{m_1/m_2}$ is the impact factor. Both the density profile of the dark matter and the fraction of particles moving more slowly than the black hole are functions of radius and time, because there is feedback from the presence of the black hole onto the spike. The initial density profile can be taken as that of Eq.~\ref{eq:profile}, and then a prescription such as that of e.g. {\fontfamily{qcr}\selectfont HaloFeedback}  \cite{Kavanagh:2020cfn} can be used to describe the evolution of the dark matter profile in the presence of the companion object in time. For intermediate and extreme mass ratio systems, orbits are in fact expected to be eccentric. This could have a non-negligible effect on the ability to measure the dark matter effect on the waveforms \cite{karydas2024sharpening,PhysRevD.105.063029}, but a full study of more accurate waveforms including post-Newtonian effects for the regime of PBHs remains to be done.

The change in evolution of the separation of the binary is equivalent to a change in frequency evolution, which enters into the evolution of the phase of the gravitational waves:
\begin{equation}
\Phi(f) = \int^{f_{\rm f}}_{f}\frac{d t}{d f^\prime}f^\prime d f^\prime.
\end{equation}
The dephasing $\delta\Phi$ is then the difference in phase between the dark matter affected system and some reference vacuum system
\begin{equation}
\delta\Phi = \Phi_{\rm V}(f_{\rm i}) - \Phi_{\rm DM}(f_{\rm i}) = 2\pi N
\end{equation}
where $f_i$ is some chosen initial frequency where the observation begins and $N$ is the number of cycles between initial and final frequencies. Naively the vacuum system would be the equivalent system with the same mass parameters, but one could also define a `best-fit' vacuum system that aims to minimise the dephasing and hence provide a conservative way of asserting the measurability of the dark matter environment \cite{Coogan:2021uqv}.

There is also an effect on the gravitational waveform due to accretion of the dark matter onto the companion black hole, but this has been shown to be less important than the effect of dynamical friction for the mass ratios of interest \cite{karydas2024sharpening}.

\subsubsection{Prospects for detection with Einstein Telescope and Cosmic Explorer}
With the improved sensitivities and wider frequency ranges of both proposed 3rd generation ground-based gravitational wave detectors Einstein Telescope and Cosmic Explorer, it could be possible to detect intermediate or extreme mass ratio PBH binaries and measure the properties of the dark matter clouds around them, if they exist. It has been shown that detecting PBH binaries in vacuum with small mass ratios will be possible, potentially using techniques borrowed from continuous wave search strategies, or hierarchical searches \cite{Magee_2018,Nitz_2021,miller2024gravitational,MILLER2021100836}. It has also been shown that if there is an undisrupted dark matter spike around the larger of the two black holes, that with one week duration signals, it will be possible to confidently measure the density of the dark matter spike to be non-zero, as well as measure the distinctive slope of the dark matter profile. For example, for a system with $m_1 = 1\,{\rm M_\odot}$ and $m_2 = 10^{-3}\,{\rm M_\odot}$, the density normalisation can be measured to precision $\rho_6 = 1.7^{+0.8}_{-0.6}\times10^{13}\,{\rm M_\odot/pc^3}$ and the slope to be $\gamma_{\rm sp}=2.22^{+0.07}_{0.05}$ where the error bars indicate the 68\% credible intervals \cite{Cole:2022ucw}. However, with a realistic mass function for the PBHs and respecting all other direct detection constraints on PBHs in the sub-solar mass range, the merger rate is expected to be low ($\sim0.3$ per year) \cite{Cole:2022ucw}, although recent studies have shown that the dark matter spike itself can have a large effect on the merger rate \cite{Jangra:2023mqp}. Future detectors will need to be lucky in order to detect these systems over their lifetimes of order years. Furthermore, various data analysis challenges such as being able to deal with these long duration signals both in terms of the search strategy and parameter estimation still need to be overcome.

\subsubsection{Parameter estimation bias between dark matter and vacuum}
\begin{figure}[b]
\sidecaption
\includegraphics[width=10cm]{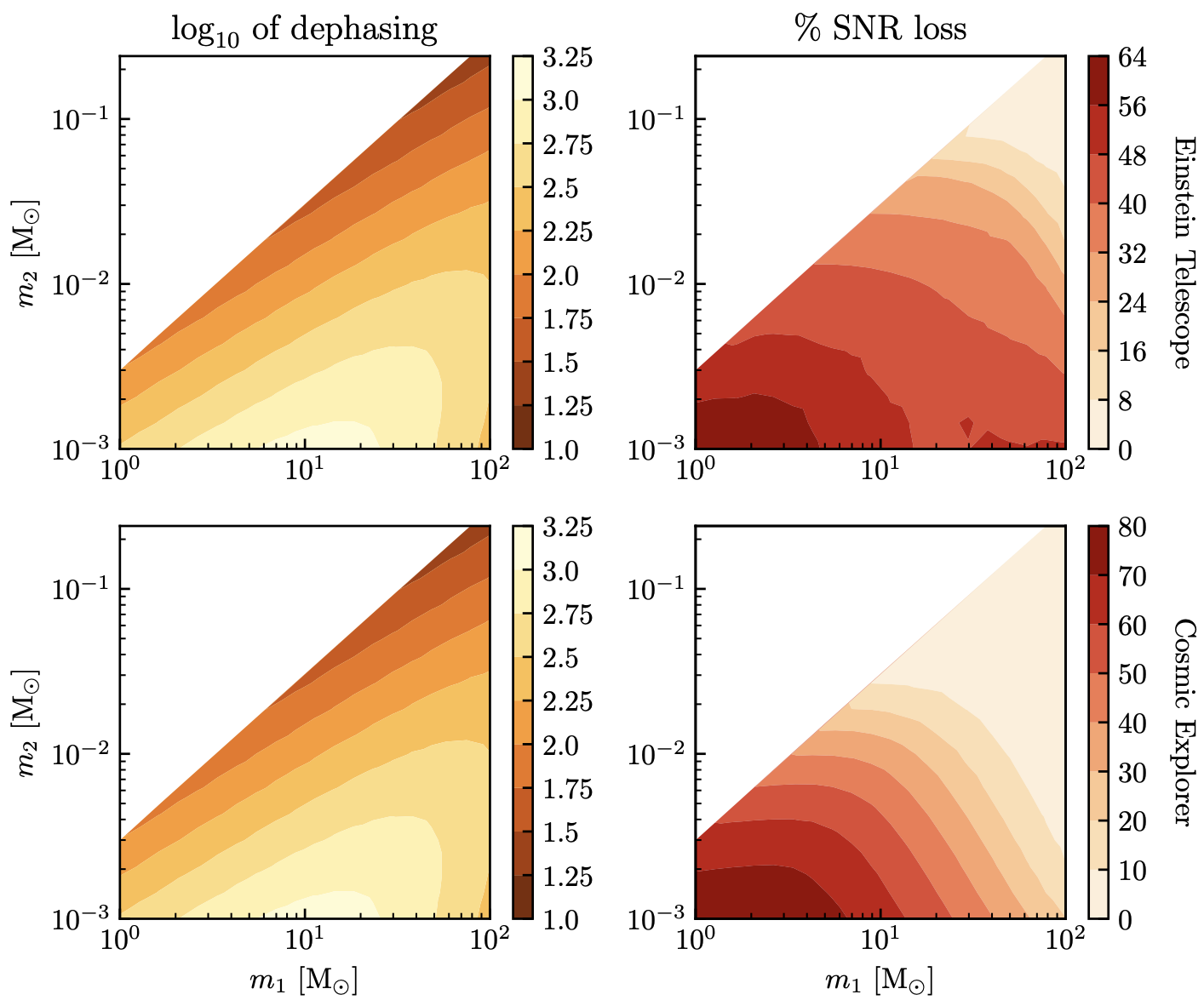}
%
%
\caption{The dephasing with respect to the best-fit vacuum system (left column) and the percentage of SNR lost with respect to the best-fit vacuum system (right column) as a function of primary $m_1$ and secondary $m_2$ mass. The top column is with respect to Einstein Telescope sensitivity and the bottom is with respect to Cosmic Explorer sensitivity. Figure reproduced with permission from \cite{Cole:2022ucw}.}
\label{fig:SNR}       
\end{figure}

It is possible, to some extent, to mimic the dark matter dephasing effect with a vacuum system of larger total mass (degeneracies with other parameters of the binary may also be possible). This degeneracy can be quantified by the amount of signal-to-noise ratio (SNR) lost in the mis-match between a dark matter gravitational waveform and its equivalent best-fit vacuum system, i.e. the vacuum template that does the best job of mimicking the dark matter-influenced system \cite{Coogan:2021uqv,Cole:2022ucw}. For a range of masses, the percentage SNR loss is shown in Fig.~\ref{fig:SNR} for both Einstein Telescope and Cosmic Explorer expected sensitivities and 1-year duration signals. 

For most of the parameter space, the SNR lost if a vacuum template is used is over 50\%. This is both promising in the sense that a full data analysis approach that accounts for the possibility of the presence of dark matter will be able to confidently distinguish it from vacuum, but also that these systems may go undetected if the analysis does not include the DM possibility in its approach. Since PBHs are expected to form dark matter haloes by necessity in this region of the parameter space where $f_{\rm PBH}=1$ is ruled out, searching for these systems in vacuum will be likely futile. However, this only applies to binaries with a very small mass ratio where the spike is expected to have survived. 

\subsubsection{Astrophysical versus primordial origin}
In the case that the primary black hole is not clearly sub-solar within the measurement errors and therefore it is ambiguous as to whether it is of astrophysical or primordial origin, the observation of the presence of a dark matter spike around the primary black hole could provide distinctive distinguishing evidence. This is due to the fact that astrophysical black holes, of masses detectable with ground-based gravitational wave detectors, are not expected to form dense spikes of dark matter around them. Even in the case of intermediate mass black holes, where it is possible to invoke the idea that they formed in isolation in the centre of deep dark matter potential wells at high redshift, the expected slope of the dark matter profile is $\gamma_{\rm sp}=7/3$ \cite{Gondolo_1999,Eda_2015}, different to that expected for primordial black holes ($\gamma_{\rm sp}=9/4$). For supermassive black holes at the centres of galaxies, it is expected that many of them will have undergone major mergers in their history, and therefore maintaining an undisrupted steep density profile is unlikely. The measurement of a dark matter profile slope of $\gamma_{\rm sp} = 9/4$ from a binary black hole inspiral would therefore be smoking gun evidence for both a primordial origin for the heavier black hole (and necessarily the lighter one since it would be sub-solar) as well as for the existence of a particle dark matter spike around it.

\subsection{Superradiance GW signatures}
As discussed in Sect.~\ref{sec:alps}, superradiance could cause a boson cloud to form around spinning primordial black holes. The most likely scenario for this combination to occur in the late universe such that the boson cloud is still present, as opposed to having extracted all of the angular momentum from the black hole and already decayed, is if the PBH is the remnant of a previous PBH merger \cite{Barausse_2009} or if a PBH is spun up significantly by accretion \cite{Luca_2020}. There are possibilities for observing the signatures of such clouds via gravitational waves, both in the case of a single PBH, in which case the cloud itself will emit gravitational waves \cite{Ghosh_2021}, or in the case of a PBH binary, in which case there could be evidence of ionisation of the cloud in the dephasing of the gravitational waveform \cite{Baumann:2022pkl,Baumann:2021fkf,Tomaselli:2023ysb}, analogous to the dynamical friction discussed in Sect.~\ref{sec:df}. In the latter case, however, the combination of requirements: that two PBHs merge to form a spinning remnant, that a boson cloud grows via superradiance, and that then a secondary binary forms with low mass ratio (such that the ionisation effect is large enough to be observed) are challenging, so we will focus on the single PBH case.

The dominant production of gravitational waves is caused by annihilation of bosons, resulting in gravitational waves with frequencies that are twice the Compton frequency \cite{Franciolini_2022}:
\begin{equation}
f\sim2\times10^6\,{\rm Hz}\left(\frac{m_b}{10^{-9}\,{\rm eV}}\right),
\end{equation}
and with an amplitude $h_0$ (in time domain hence dimensionless) of
\begin{equation}
h_0 \sim 5 \times 10^{-25}\frac{1}{l}\left(\frac{\alpha}{0.1}\right)\left(\frac{\epsilon}{10^{-3}}\right)\left(\frac{d_L}{\rm {kpc}}\right)^{-1}\left(\frac{m_b}{10^{-9}\,{\rm eV}}\right)^{-1}
\end{equation}
where $\alpha=Gm_{\rm PBH}m_b$ is the gravitational fine structure constant, $l$ is the orbital angular momentum number of the bosons and $\epsilon$ is the fraction of the PBH mass which is transferred to the cloud. For superradiance to take place, $\alpha/l<0.5$.
Therefore, for example, a spinning PBH with mass $m_{\rm PBH}=1\,{\rm M_\odot}$ could form a cloud of bosons with masses $m_b=10^{-10}\,{\rm eV}$ via superradiance, which would produce monochromatic gravitational waves due to their self-annihilation at a frequency of $f=2\times10^5\,{\rm Hz}$ with maximum amplitude $h_0\sim2.5\times10^{-23}$, assuming that $\alpha/l=0.5$, $\epsilon=10^{-3}$ and $d_L=1\,{\rm kpc}$. 

Current ground-based gravitational wave detectors have bandwidth up to a maximum frequency of approximately $10^4\,{\rm Hz}$, and therefore would only be sensitive to bosons of masses $m_b<10^{-11}\,{\rm eV}$, which would form clouds around spinning black holes of masses larger than $m_{\rm PBH}>10\,{\rm M_\odot}$. For these masses, it would be difficult to be certain that the black hole around which the boson cloud has formed is primordial. Therefore, it will be necessary to go to the higher frequency regime, where data from various experiments such as ADMX \cite{Khatiwada_2021} can be re-purposed for high-frequency gravitational wave searches \cite{Domcke_2022}. According to recent estimates, the current sensitivities are not good enough to detect high frequency gravitational waves from the annihilation of bosons and therefore technological improvements or dedicated search strategies in this frequency regime will be required if there is hope for this channel of detection \cite{Franciolini_2022}.

\section{Dark matter particles from PBH evaporation}\label{sec:evap}
Primordial black holes evaporate due to Hawking radiation \cite{Hawking:1975vcx}, and those lighter than $\mathcal{O}(10^{-18})\,{\rm M_\odot}$ will have completely evaporated before today. There are strong bounds on the abundance of primordial black holes in the mass range $10^{-24}-10^{-16}\,{\rm M_\odot}$ due to the non-observation of the expected effects of their evaporation from big-bang nucleosynthesis (BBN), the extra-galactic gamma-ray background, and the cosmic microwave background (CMB) \cite{Acharya_2020,AUFFINGER2023104040,Keith_2020,Boudaud_2019}. However, black holes with masses less than $\sim10^{-24}\,{\rm M_\odot}$ could have produced dark matter particles during their evaporation before BBN \cite{PhysRevD.108.015005}. 


PBHs produce all particles with masses below the Hawking temperature $T_{\rm BH}$ of the black hole:
\begin{equation}
k_BT_{\rm BH} = \frac{1}{8\pi}\frac{(M_Pc^2)^2}{M_{\rm BH}c^2}
\end{equation}
where $k_B$ is the Boltzmann constant and $M_P$ is the Planck mass. There are two possibilities for producing stable dark matter particles from a single PBH. Either dark matter particles are emitted immediately after the formation of the PBH with number
\begin{equation}
N_X = \frac{g_{X,H}}{g_{*,H}}\frac{4\pi}{3}\left(\frac{M_{\rm BH}}{M_P}\right)^2
\end{equation}
that depends only on the mass of the PBH (at formation), and $g_{X,H}$, $g_{*,H}$  are the number of stable dark matter particles with masses below $k_{\rm B}T_{\rm BH}$ and total number of particles with masses below $k_{\rm B}T_{\rm BH}$ respectively. Or, dark matter particles are emitted towards the end of the lifetime of the PBH, once the temperature of the PBH has increased enough to be larger than the mass of the heavy particles with number
\begin{equation}
N_X = \frac{g_{X,H}}{g_{*,H}}\frac{1}{48\pi}\left(\frac{M_P}{M_{\rm X}}\right)^2
\end{equation}
that only depends on the mass of the dark matter particle, $M_{\rm X}$ \cite{Masina:2020xhk}. There are strong constraints on such particles from requiring them to be sufficiently cold so as not to conflict with structure formation observations. However there may be ways to relax these constraints by considering a broad, and more realistic, PBH mass distribution \cite{chattopadhyay2023dark} or to allow for entropy non-conservation \cite{Fujita_2014}. Furthermore, there are interesting phenomenological signatures to be explored even if the entire dark matter abundance is not radiated by PBHs. For example, if PBHs are abundant enough to trigger an early-matter dominated era \cite{Hooper_2019}, and they produce axions as they evaporate, then the temperature at which those axions will start to oscillate will be altered \cite{Bernal_2021}.

\section{Conclusion}
Primordial black holes may co-exist with a population of dark matter particles if they do not make up all of the dark matter themselves. There are interesting prospects for detection or constraints on both species due to interactions between the two. For example, weakly interacting massive particles would form dense haloes around PBHs, which on certain mass ranges, would have induced a signature in the diffuse gamma-ray background which has not been detected, thus placing constraints on large portions of the parameter space for both species if they are to co-exist. Axions or axion-like particles could also form dense haloes around PBHs which may be detectable via dephasing signatures in the gravitational waveform of PBH binaries. Or, if there exists a population of spinning primordial black holes along with a population of axion-like particles, then a boson cloud could form around the PBH due to superradiance. This system requires a combination of initial conditions to align, but would have very distinctive signatures in gravitational wave observations. 

Alternatively, instead of co-existing, PBHs may have been the predecessors of our current dark matter population. Dark matter particles could have been produced in the early universe, before BBN, due to the evaporation of very light primordial black holes. Constraints on how warm dark matter can be due to structure formation observations are perhaps the most tightly constraining on this scenario, but ways of relaxing or alleviating these constraints have been considered.

Given that all but one window of the PBH mass range has been constrained such that the totality of the dark matter energy density budget cannot be explained by primordial black holes, i.e. $f_{\rm PBH}<1$, it is important to consider a combined scenario for dark matter if PBHs are to play a role. Observations from the CMB, BBN and the gamma-ray background have already provided additional constraints on a sub-population of PBHs combined with WIMPs, and there are prospects for gravitational wave observations in particular to provide new constraints on distinctive properties of a multi-component model of dark matter that involves primordial black holes.

 \bibliographystyle{unsrt}
 \bibliography{chapter_15.bib}

\end{document}